\documentclass[aps,pre,onecolumn,floatfix,amsmath,amssymb]{revtex4}

\usepackage{graphicx}
\usepackage{amsmath}

\begin{document}

\title{Stochastic population dynamics in turbulent fields}
\author{M.~H. Vainstein}
\affiliation{Instituto de F\'{\i}sica and Centro Internacional de F\'{\i}sica da Mat\'eria Condensada, Universidade de Bras\'{\i}lia, CP 04513, 70919-970, Bras\'{\i}lia-DF, Brazil}

\author{J.~M. Rub\'{\i}}
\email[Corresponding author: ]{mrubi@ub.edu}
\affiliation{Departament de F\'{\i}sica Fonamental, Universitat de Barcelona, Diagonal 647, 08028 Barcelona, Spain}

\author{J.~M.~G. Vilar}
\affiliation{Integrative Biological Modeling Laboratory, Computational Biology Program, Memorial Sloan-Kettering Cancer Center, 1275 York Avenue, Box 460, New York, NY 10021, USA}


\begin{abstract}
The behavior of interacting populations typically displays irregular temporal and spatial patterns that are difficult to reconcile with an underlying deterministic dynamics. A classical example is the heterogeneous distribution of plankton communities, which has been observed to be patchy over a wide range of spatial and temporal scales. Here, we use plankton communities as prototype systems to present theoretical approaches for the analysis of the combined effects of turbulent advection and stochastic growth in the spatiotemporal dynamics of the population. Incorporation of these two factors into mathematical models brings an extra level of realism to the description and leads to better agreement with experimental data than that of previously proposed models based on reaction-diffusion equations.
\end{abstract}

\maketitle

\section{Introduction}
\label{sec:intro}

Plankton patchiness has many causes, with origins both in biological and physical factors. Over small scales ($1\,$mm to $10\,$m), biological factors on the individual scale such as mating, predator avoidance, finding food and the diel vertical migration (upward and downward swimming at certain times during the $24\,$h day) are crucial factors for the emergence of plankton patchiness~\cite{Folt99}. On larger length scales ($10\,$m to $100\,$km), physical processes such as turbulence, currents and eddies are the principal causes of patterns. 

It has been shown that the relative intensity of zooplankton patchiness is greater than that of phytoplankton at all spatial scales~\cite{Mackas77}.
 Mackas and Boyd~\cite{Mackas79} developed a method to count individual particles in a continuous stream of seawater. This permitted them to use shipborne sampling to find transects of zooplankton abundance and chlorophyll fluorescence, which provides an appropriate method for estimating phytoplankton concentration, and to make a spectral analysis of the spatial heterogeneity.  They found that small spatial scale contribution to the total patchiness is much greater for zooplankton, which is reflected in the fact that power spectra are less steeply sloped for zooplankton than for phytoplankton. Moreover, they found that the spatial patterns of phytoplankton and zooplankton are negatively correlated.

Satellite imagery allows the obtention of greater spatial coverage and a better time resolution than shipborne sampling. The analysis of changes in the spectrum of visible light due to absorption and fluorescence of chlorophyll pigments allow the estimation of phytoplankton biomass. Gower \emph{et al.}~\cite{Gower80} performed a spectral analysis of satellite images, the results of which lead them to believe that phytoplankton patchiness is controlled by mesoscale ($10$-$100\,$km) water motions. Phytoplankton have small mobility, consisting of vertical migration limited to over a few meters per day and may therefore be considered almost as a passive scalar advected by the ocean currents.  

Many model systems have been put forward to explain pattern formation in such planktonic systems. To cite a few, Abraham~\cite{Abraham98} has considered the role of non-diffusive advection in plankton pattern formation and L\'opez {\emph et al.}~\cite{Lopez01} and Hern\'andez-Garcia {\emph et al.}~\cite{Hernandez-Garcia02} offer a comprehensive overview of modeling of plankton dynamics as chaotic advection.  
   However, the important effects of noise have usually been neglected.
 
This work is organized as follows: in the next section, we briefly discuss models of population growth used in ecology; in Sect.~\ref{sec:noise} we describe the effects of noise in such models; in Sect.~\ref{sec:space} we analyze how the inclusion of diffusion and advection can generate patterns in population dynamics; in Sect.~\ref{sec:spectral}, we discuss spectral analysis, one of the main tools used in studying plankton patterns; we give final remarks in Sect.~\ref{sec:conclusion}. 
\begin{figure}[t]
\begin{center}
\resizebox{0.75\columnwidth}{!}{
\includegraphics[height=4cm]{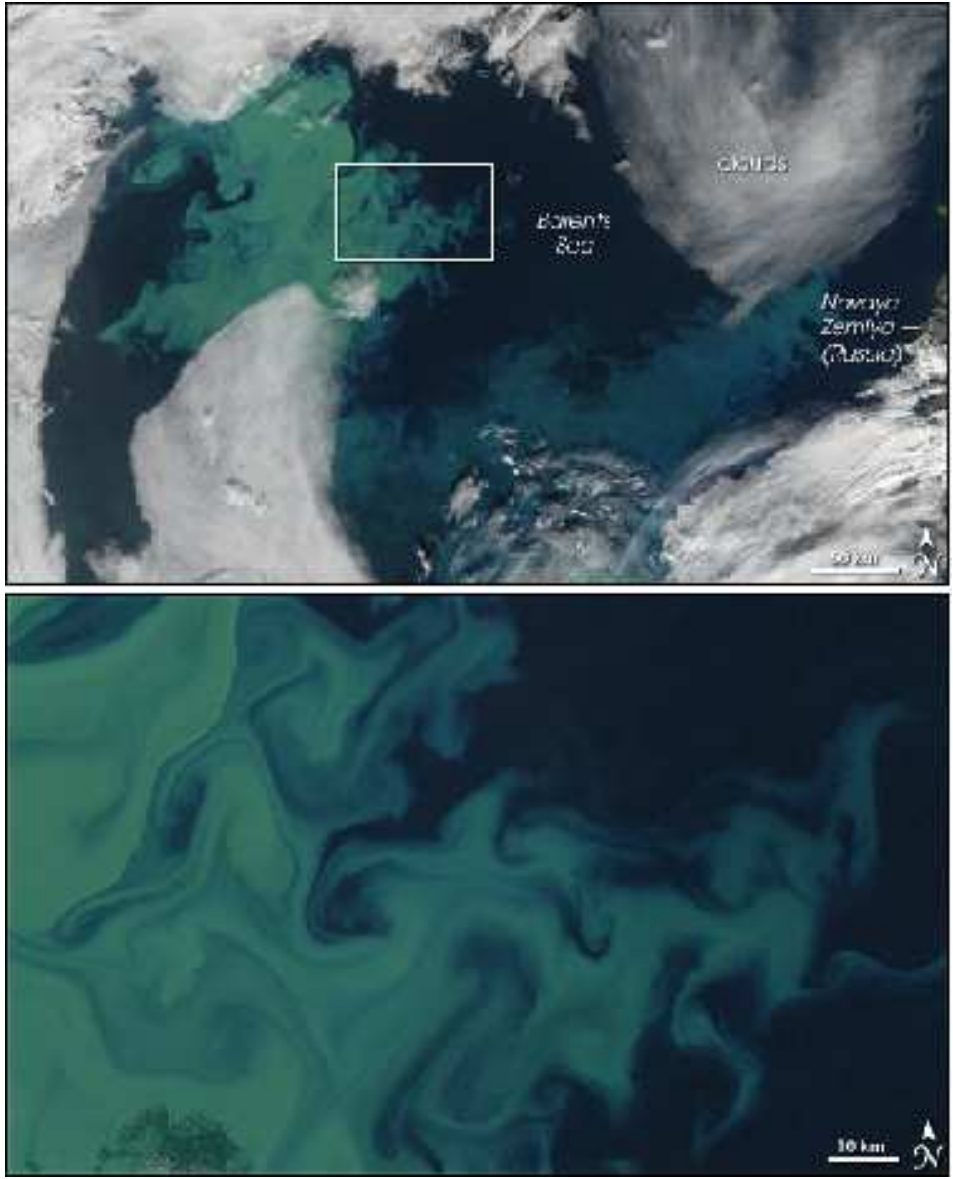}
}
\end{center}
\caption{Image of a phytoplankton  bloom in the Barents Sea north of Russia, captured by the Moderate Resolution Imaging Spectroradiometer (MODIS) on NASA's Aqua satellite on August 29, 2006. Source: NASA's Earth Observatory~\cite{NASA}.}
\label{fig:satellite}  
\end{figure}

\section{Population growth models}
\label{sec:eco_model}
In ecology, continuous growth population dynamics is usually modelled by a system of differential equations which correspond to the conservation equations for each population, in which the rate of change of the different species is related to birth, death and migration processes. Predator-prey models are usually defined through
\begin{align}
\frac{d N}{d t}&= F_N(N,P), \text{ and} \\
\frac{d P}{d t}&= F_P(N,P), 
\end{align}
where $N(t)$ represents the prey population, $P(t)$ the predator population and the functions $F_N(N,P)$ and $ F_P(N,P)$ account for the usually nonlinear interactions between the species.  A general starting point is the Lotka-Volterra model, in which $F_N(N,P) = N(a-bP)$, and $F_P(N,P)=P(cN-d)$, with $a$, $b$, $c$ and $d$ positive constants. However, this model has some drawbacks, since its solutions are not structurally stable~\cite{Murray03}, meaning that small perturbations can exert large effects on the amplitude of oscillation, for example. An improvement is given by a logistic growth of the prey, which limits the prey population even in the absence of predators. This feature is taken into account by the function $F_N(N,P) = rN(1-N/K)-cPf(N)$, where $r$ is a linear birth rate, $K$ is the carrying capacity of the environment, and $c$ a parameter related to the predation rate. The response function $f(N)$, to be more realistic, should saturate to account for satiation. A possible choice is the Holling type III functional response $f(N)=N^2/(1+N^2)$~\cite{Holling59}. An equation with these terms and considering a constant number of predators $P(t)\equiv P$ was used to model  the outbreak of the spruce budworm~\cite{May77}. In that case, the carrying capacity is related to the foliage (food) available in the trees, and the functional response models a switching on of the predation rate by birds at a given threshold population of the prey. In other words, if the prey population is low, the predator (birds) will find a different source of food and if the prey population is high enough, it will become a source of food. Moreover, the saturation in the functional response represents the fact that there is a maximum uptake of food by the birds.

A model similar to the one for the spruce budworm was used together with a time-dependent growth rate  $r\equiv r(t)$, and $F_P(N,P)=P(gf(N)-\epsilon)$, with $g$ and $\epsilon$ positive constants, to model the interaction of phytoplankton and zooplankton with different timescales  that lead to rapid increases of phytoplankton population known as ``spring blooms'' and ``red tides''~\cite{Truscott94}.

\section{The role of noise}
\label{sec:noise}
\begin{figure}[ht]
\begin{center}
\resizebox{0.75\columnwidth}{!}{
\includegraphics[angle=270,width=4cm]{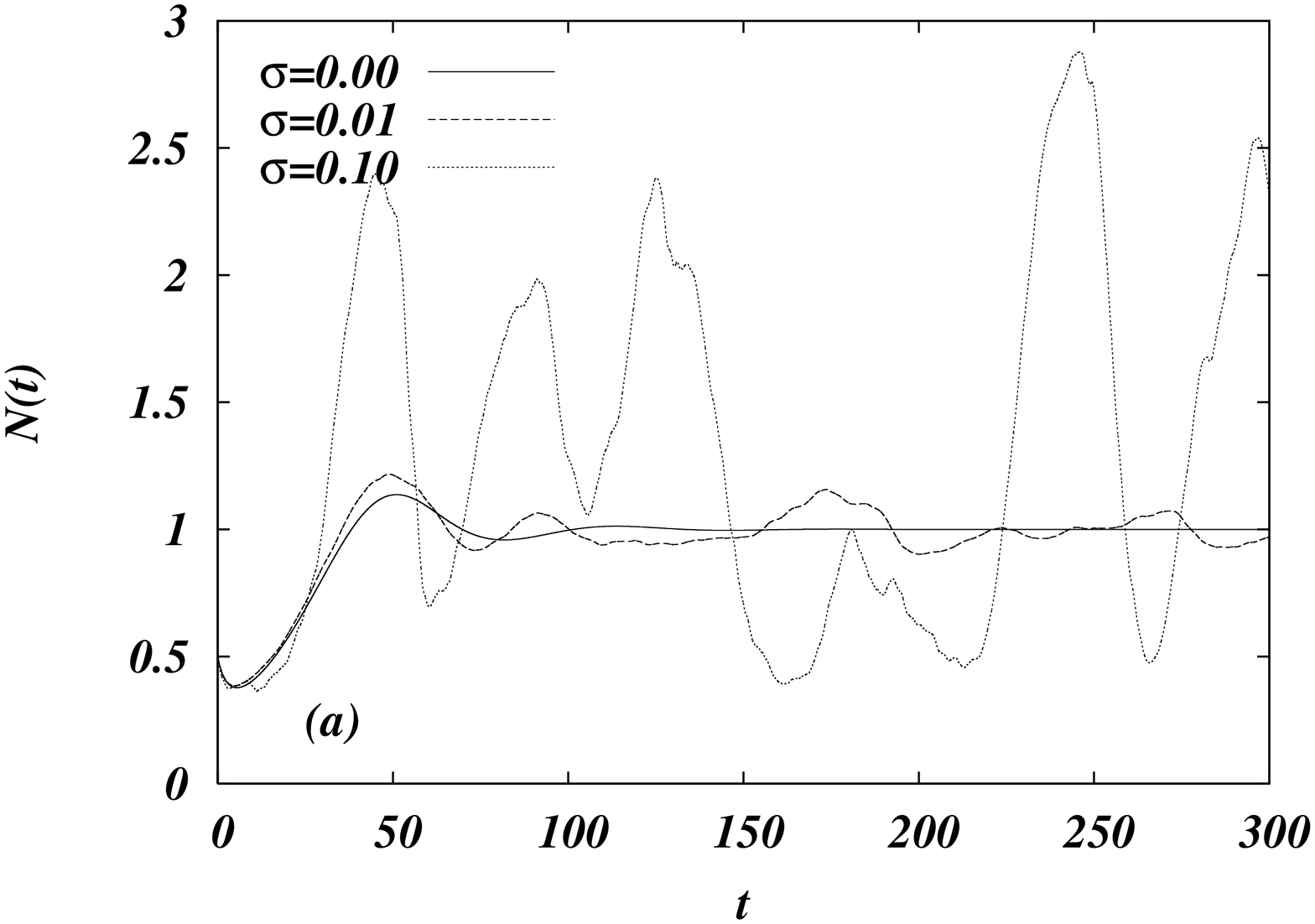}
\includegraphics[angle=270,width=4cm]{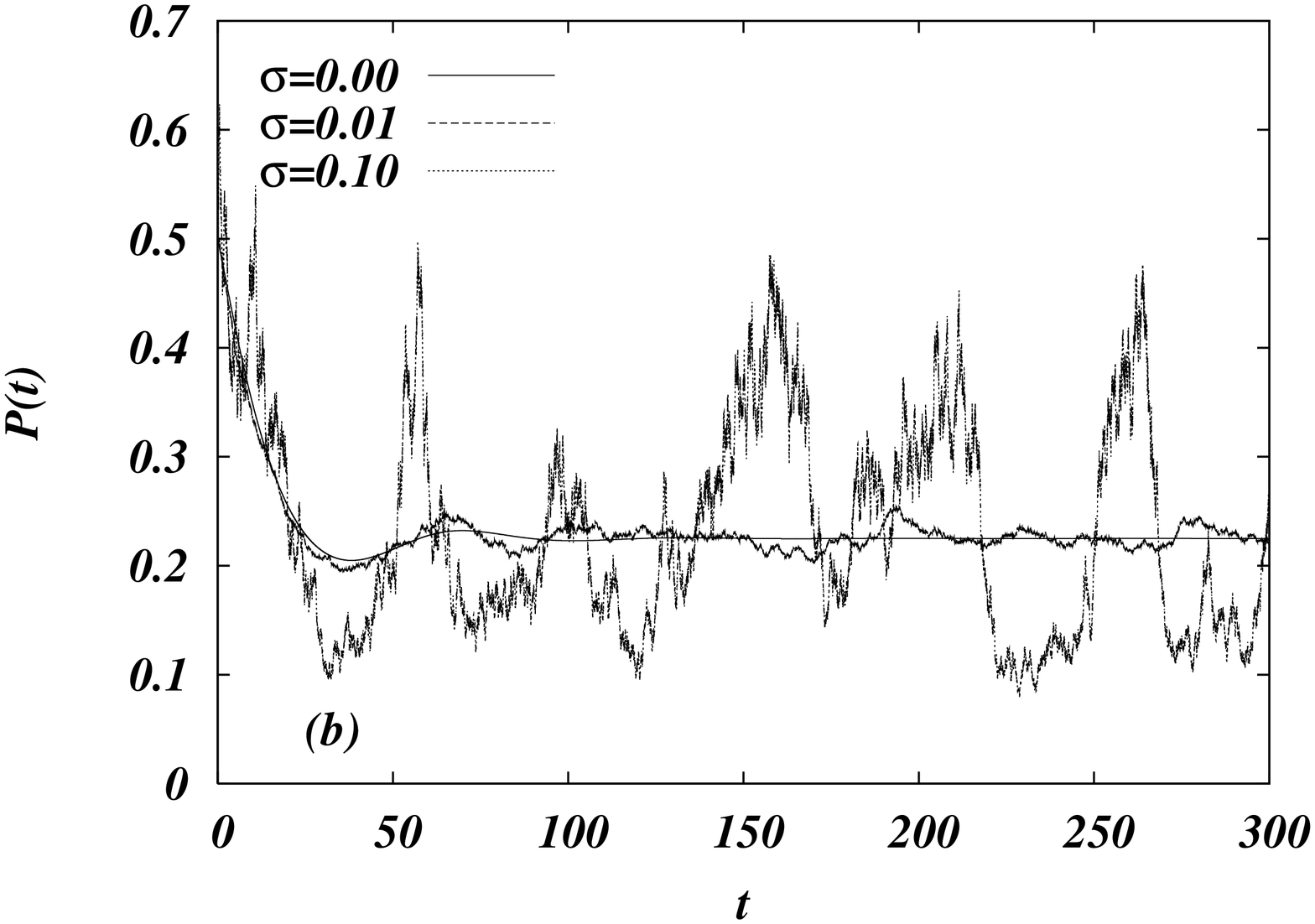}
}
\end{center}
\caption{Population dynamics given by Eqs.~(\ref{phyto_dyn}) and (\ref{zoo_dyn}) with  $r=0.3$, $K=4.0$, $c=2.0$, $g=0.1$, $\epsilon=0.05$. The populations fluctuate around the stable equilibrium $N_{eq}=1.0$ and $P_{eq}=0.225$. For larger values of the noise, the system can be driven to the unstable equilibrium in which the zooplankton population becomes extinct. (a) Phytoplankton concentration. (b) Zooplankton concentration.}
\label{fig:noise}  
\end{figure}

Deterministic population models rarely capture all of the features of dynamical systems, since real systems are influenced by changes in the environment which are in essence random. This inherent unpredictability of the environment may manifest itself in fluctuating birth rates, carrying capacities and other parameters which characterize biological systems. A deterministic treatment is more adequate for very large populations. However, the population of zooplankton is considerably smaller than that of phytoplankton and a stochastic approach is more realistic. The reason for the inclusion of noise is the fact that zooplankton interact with fish and whales which are present in even smaller numbers and are far from being evenly distributed.

Nevertheless, in many cases, the inclusion of noise may destabilize a system and lead to the extinction of a given species. The linear stability of a system of $m$ species
\begin{equation}
\frac{d N_i}{d t}= F_i(N_1(t),\dots,N_m(t))
\end{equation}
depends on the eigenvalues of the matrix with components $a_{ij}\equiv (\partial F_i /\partial N_j)_{eq}$ which governs the dynamics near equilibrium in the linear approximation. For a system without noise to be stable, it is necessary that all eigenvalues have negative real parts. If we define  $\Lambda$ as minus the largest real part of the eigenvalues, then the stability criterion becomes $\Lambda >0$. In the presence of fluctuations characterized by variance $\sigma^2$, the stability criterion should be changed to $\Lambda > \sigma^2$~\cite{May01}. The noise will generally tend to decrease the average population number by increasing the severity of the population fluctuations. To see the effects of noise, we will consider the model 
\begin{align}
\frac{d N}{d t}&= rN(1-N/K) - c P N^2/(1+N^2), \label{phyto_dyn}\\
\frac{d P}{d t}&= P(g  N^2/(1+N^2) -\epsilon) + \xi(t), \label{zoo_dyn}
\end{align}
where $\xi(t)$ is a random term, and  $N$ and $P$ are the phytoplankton and zooplankton populations, respectively . As pointed out in~\cite{Mackas79}, the intrinsic rate of zooplankton growth is lower than that of phytoplankton, what should lead to the formation and maintenance of only very large scale features. They concluded that a mechanism (maybe behavioral in origin) with shorter time scale should be at play to account for the small-scale patchiness observed. This is the motivation for the inclusion of a multiplicative noise term,  since fluctuations originate from  processes which depend on the local concentration, such as the reproduction process and the consumption by larger animals. Therefore, we take  $\langle \xi(t) \xi(t')\rangle=2[\sigma P(t)]^2 \delta (t-t')$, where $\sigma P(t)$ is a fluctuating growth rate. An \emph{a posteriori} reason for not considering additive noise is that it may allow the concentration of zooplankton to reach negative values, which is unreasonable. In the study of stochastic population dynamics in the Poisson approximation, an expression for the fluctuations around the deterministic limit in which the noise amplitude is state dependent is also arrived at~\cite{Aparicio01,Solari03}. 

The inclusion of noise of this type may easily destabilize the system, since unrealistically large fluctuations may arise even for very small values of $\sigma$. In the absence of noise, the populations will converge to the stable equilibrium given by
\begin{align}
N_{eq}&=\sqrt{\frac{\epsilon}{g-\epsilon}} \label{phyto_eq} \\
P_{eq}&=\frac{gr}{c\epsilon} \left[ \frac{K\sqrt{\epsilon (g-\epsilon)}-\epsilon}{K(g-\epsilon)}\right]. \label{zoo_eq}
\end{align}
  Figure~\ref{fig:noise} shows how the populations fluctuate around their average stable values $N_{eq}$ and $P_{eq}$, but the fluctuations can eventually take the system to the unstable equilibrium $N=K$ and $P=0.0$. This negative effect of noise can be attenuated by the inclusion of space variables and diffusion or advection, as will be seen in the next section.

The inclusion of noise does not always have a negative impact on dynamics; in some systems, its  effects can be particularly important and can lead to unexpected results such as stochastic resonance~\cite{Benzi81,Gammaitoni98}, a phenomenon characterized by the enhancement of the response of a system to a periodic driving force in the presence of noise. Stochastic resonance has been studied in many systems, including spatially extended systems and pattern-forming systems
, such as the Swift-Hohenberg equation~\cite{Vilar97}. Another interesting possibility is the suppression of internal noise by the application of an external noise source~\cite{Vilar01}. 

\section{The inclusion of space variables}
\label{sec:space}
The models considered up to now are in a sense incomplete, since they provide a mean-field description of the variations of the population in which only global changes are considered.  Even though average population densities and even some naturally occurring phenomena such as outbreaks can be predicted, many other features of the systems are left out, pattern formation being one of these. 

\subsection{Diffusion}

Turing~\cite{Turing52} studied how diffusion together with nonlinear local interactions could lead to the emergence of heterogeneities even when starting from homogeneous conditions in a homogeneous environment. Reaction-diffusion equations of the form 
\begin{equation}
\frac{\partial \mathbf{u}}{\partial t} = \mathbf{f}(\mathbf{u})+D\nabla^2 \mathbf{u}
\end{equation}
 have been used since then to model pattern formation in system on a wide range, from chemical reactions to bacterial chemotaxis and animal coat patterns~\cite{Murray03,Cross93}.
Initially, models of plankton spatial heterogeneity and patchiness were based on these types of equations, in which the populations are investigated theoretically by some variation on the predator-prey models derived from on the  Lotka-Volterra model with an added diffusive term. A study of the plankton population stability with this model was performed by Steele~\cite{Steele74}. 
Levin and Segel~\cite{Levin76} introduced an autocatalytic effect in phytoplankton density and differential dispersal rates  which favor higher herbivore motility to account for the origin of planktonic patchiness. With their hypothesis, they found a transition from a uniform stable state to a new steady state in which plant and herbivore are more concentrated in certain regions, depending on the model parameters. 
However, in the case of planktonic populations, diffusion is only important for the movement of plankton on small scales (centimeter scales), in which many biological factors are also important. 
 Moreover, the experimental results from~\cite{Mackas79} show that zooplankton is more patchily distributed and indicate that analytical models used for describing phytoplankton which are based on a perturbative analysis of the scale-dependent balance between turbulent diffusive flux and exponential reproductive growth are inapplicable for explaining the zooplankton heterogeneity over these scales. In simulations with models based on reaction-diffusion equations, zooplankton is found to be less patchily distributed than phytoplankton, contrary to experimental observations~\cite{Levin76,Levin92}. Also, satellite imagery (see Fig.~\ref{fig:satellite}) has displayed eddy patterns that represent the phytoplankton biomass and thus demonstrated that plankton patterns in the ocean occur on much broader scales and therefore mechanisms other than diffusion should be considered. 
\begin{figure}[t]
\begin{center}
\resizebox{0.75\columnwidth}{!}{
\includegraphics[angle=270]{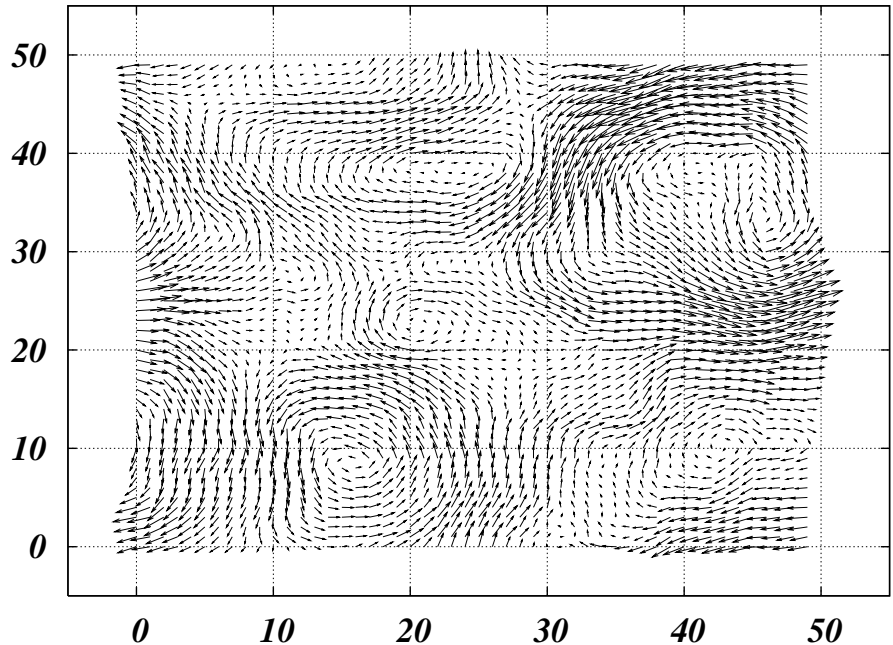}
}
\end{center}
\caption{Velocity distribution for a $50 \times 50$ mesh with spacing of $0.25\,$km, obtained from the seeded eddy model with $12000$ eddies distributed randomly with $R_{min}=1$, $R_{max}=16$ and $a=0.1\,\text{d}^{-1}$. The  arrows indicate the velocity field in each lattice point. }
\label{fig:velocity}  
\end{figure}

\subsection{Advection}
\begin{figure}[h]
\begin{center}
\resizebox{0.75\columnwidth}{!}{
\includegraphics[angle=270]{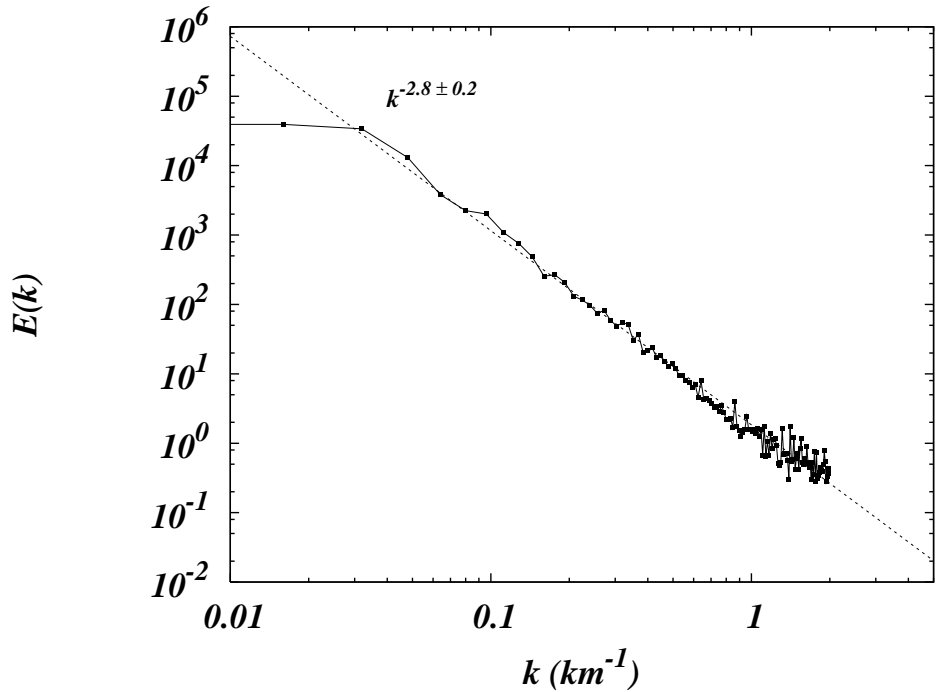}
}
\end{center}
\caption{Energy spectrum of a $250 \times 250$ with $80000$ eddies with $1<R_i<64$. The fit for larger wavenumbers is in good agreement with that expected for geostrophic turbulence ($k^{-3}$). }
\label{fig:velocity_spectrum}  
\end{figure}

Experiments performed on large scale in the ocean and satellite imagery have replaced the concept of slow ocean currents by the concept of a continuous distribution of more energetic eddies, which may have space scales up to the order of $100$km. In these larger scales, the dominant form of motions is due to turbulent lateral diffusion generated by strong ocean currents and as such is better modelled by an advective term rather than by diffusion, as considered in the previous section. However, turbulent motion is computationally expensive and therefore can not be easily included in models of population dynamics. This difficulty can be circumvented by the use of a simpler flow model such as the two-dimensional non-divergent seeded eddy model~\cite{Dyke85} to simulate geostrophic turbulence, which has an energy spectrum $E(k)\propto k^{-3}$ as proposed by Kraichnan and Charney~\cite{Kraichnan67,Charney71}.  The stream function $\psi(\mathbf{r})$ in this model is given by  
\begin{equation}
\psi(\mathbf{r})=a\sum_{i=1}^{n}(\pm)_i R_{i}^2 e^{-(\mathbf{r}-\mathbf{r}_i)^2/2R_{i}^2},
\end{equation}
 where $R_i$ and $r_i$ are the radius and position of each eddy and $a$ is a calibration constant. The velocity field is then given by 
\begin{equation}
\mathbf{v}\equiv\mathbf{v}(x,y)=\left( -\frac{\partial \psi}{\partial y},\frac{\partial \psi}{\partial x} \right) .
\end{equation}

The probability distribution for a radius size $R$ is taken to be $p(R)\propto R^{-3}$, with lower radius $R_{min}$ and higher radius $R_{max}$,  and the signs are taken arbitrarily for each eddy. A typical velocity distribution generated by this model is seen in Fig.~\ref{fig:velocity}. The energy power spectrum, which approximates that expected for geostrophic turbulence is displayed in Fig.~\ref{fig:velocity_spectrum} (for further discussion, see Sect.~\ref{sec:spectral}).

 Abraham~\cite{Abraham98} introduced non-diffusive advection to generate patchiness and found that the characteristic spatial patterns of phytoplankton and zooplankton are a consequence of the timescales of their response to changes in their environment caused by turbulent advection.  He extends a model based on logistic growth~\cite{Levin76} for phytoplankton by introducing  a maturation time for zooplankton and shows that this maturation time is responsible for the determination of spatial structure. However, in his model, he does not include grazing saturation and the type of maturation time introduced can lead to non-realistic situations, such as growth of zooplankton in the absence of phytoplankton.  His biological model consists of three coupled differential equations for the carrying capacity (maximum phytoplankton concentration attainable), for the phytoplankton and for the zooplankton. The carrying capacity continuously relaxes towards a spatially varying background value. He finds that different distributions are due to different response rates to changes in the environment caused by turbulent advection. Neufeld \emph{et al.}~\cite{Neufeld02} modelled phytoplankton blooms which occur both naturally as a result of seasonal changes and as the result of ocean fertilization experiments using similar equations, with terms for logistic growth of phytoplankton, grazing by zooplankton and  growth and mortality for zooplankton.

\begin{figure}[t]
\begin{center}
\resizebox{0.75\columnwidth}{!}{
\includegraphics[angle=270]{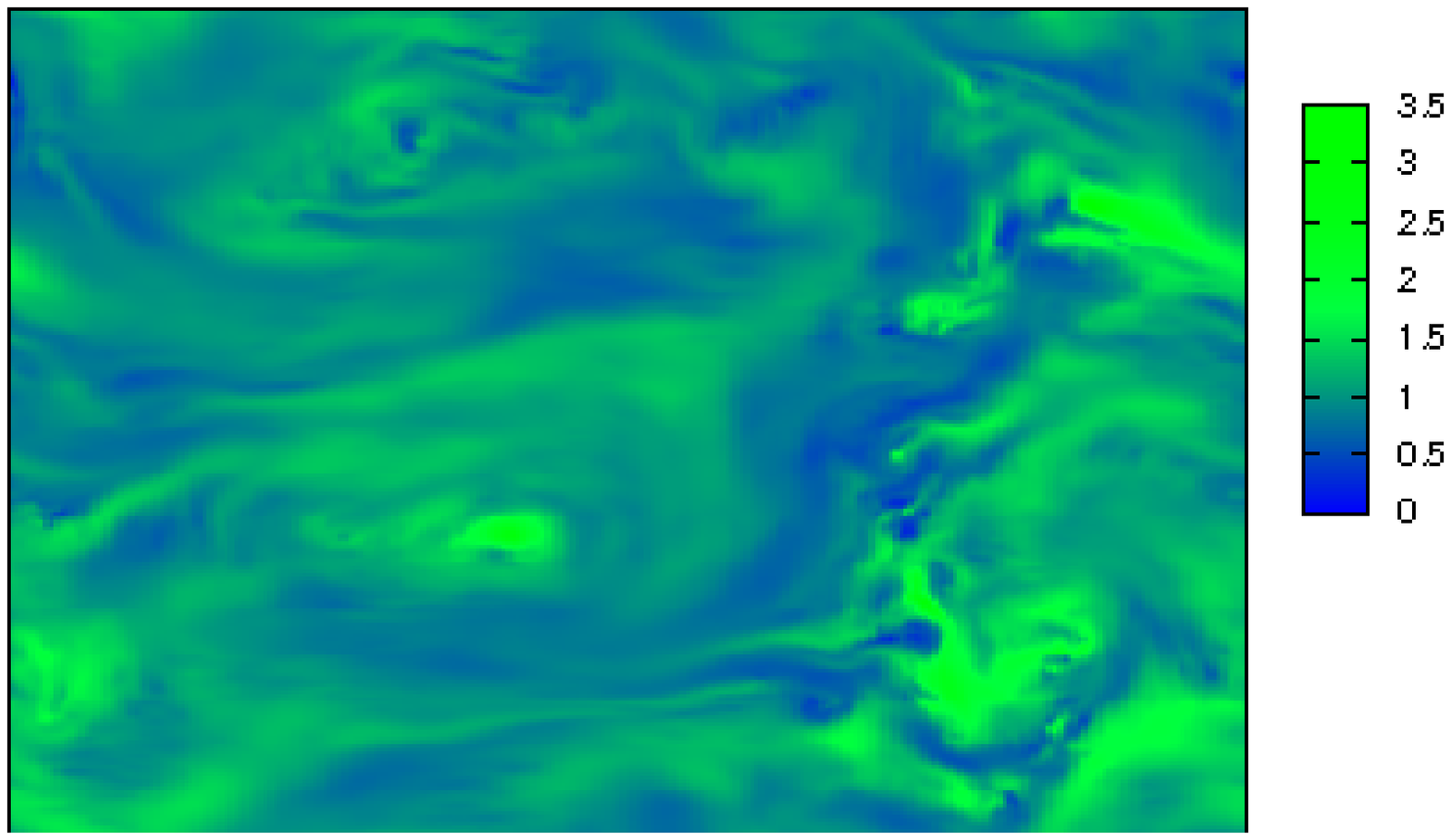}}\\
\resizebox{0.75\columnwidth}{!}{
\includegraphics[angle=270]{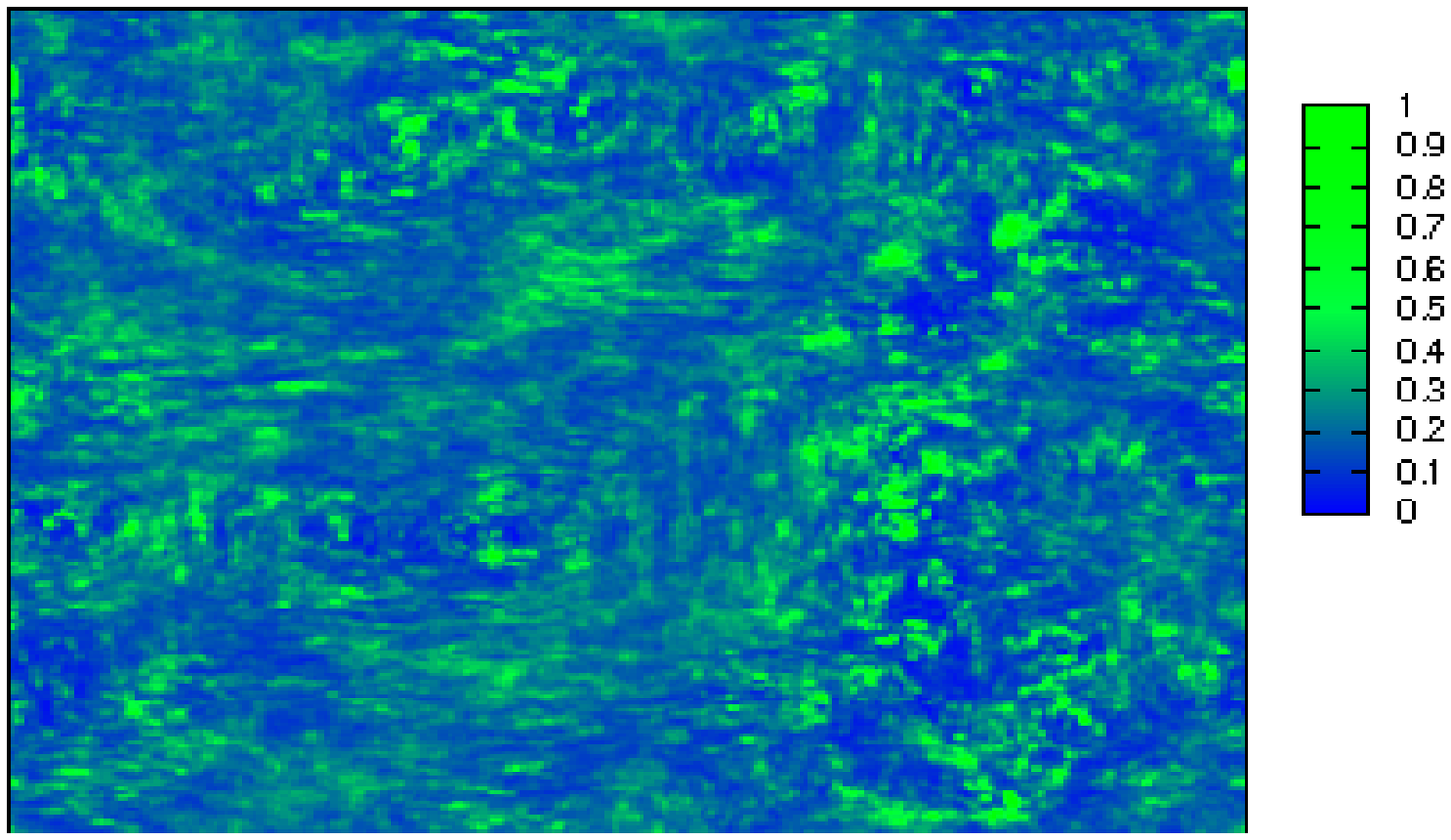}
}
\end{center}
\caption{Phytoplankton (top) and zooplankton (bottom) distribution. The phytoplankton distribution follows more closely a passive tracer distribution (not shown). However, contrary to a passive tracer pattern, the plankton patterns do not remain stationary after a transient period. Results from the simulation of Eqs.~(\ref{fevol}) and (\ref{zevol}) on a $250 \times 250$ lattice corresponding to an area of $62.5 \times 62.5\, \text{km}^2$. The values of the parameters are  $\sigma=1.0$ and the others are the same as in Fig.~\ref{fig:noise}.
}
\label{fig:pattern}      
\end{figure}

In the model we consider~\cite{Vilar03}, we include both a noise term described in Sect.~\ref{sec:noise} and turbulent advection.  It corresponds to the equations 
\begin{align}
\frac{\partial N}{\partial t}&= F_N(N,P)-\mathbf{v}\cdot \nabla N, \text{ and} \label{fevol} \\
\frac{\partial P}{\partial t}&= F_P(N,P)-\mathbf{v}\cdot \nabla P + \xi(\mathbf{r},t),
\label{zevol}
\end{align}
where the noise term is characterized by $\langle \xi(\mathbf{r},t) \xi(\mathbf{r}',t')\rangle=2[\sigma P(\mathbf{r},t)]^2\delta (\mathbf{r}-\mathbf{r}')\delta (t-t')$,  $F_N(N,P)$ and $F_P(N,P)$ are the same as in Eqs.~(\ref{phyto_dyn}) and (\ref{zoo_dyn}).  We focus on the intrinsic dynamics of the interactions between populations, which can be approximated as local in both space and time. Therefore, we assume that the noise is uncorrelated. Colored noise in space and time would arise if we considered external perturbations, such as temperature fluctuations. We denote by $N(\mathbf{r},t)$ and $P(\mathbf{r},t)$ the phytoplankton and zooplankton concentrations, respectively.  The equations are discretized on a two-dimensional mesh with periodic boundary conditions and solved with a fourth order Runge-Kutta method~\cite{Press02} for the deterministic phytoplankton growth and with the Milstein scheme for stochastic differential equations for the zooplankton growth~\cite{Kloeden95}. The advective term was solved using the second upwind scheme or donor cell scheme~\cite{Chung02}, a method used in  problems of computational fluid dynamics. 
 The initial condition is given by homogeneous populations of zooplankton and phytoplankton at their stable equilibrium which corresponds to $N_{eq}=1.0$ and $P_{eq}=0.225$ for the parameter values as in Fig.~\ref{fig:noise}.

\begin{figure}[t]
\begin{center}
\resizebox{0.75\columnwidth}{!}{
 \includegraphics[angle=270,totalheight=3cm]{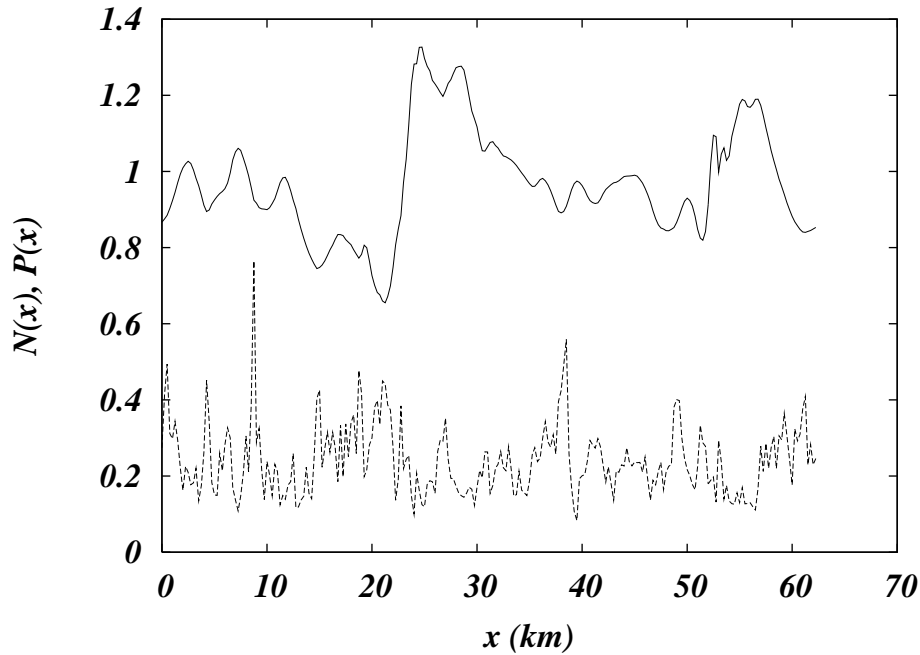} }
\end{center}
\caption{Transect of the system showing the variations in plankton concentration. Zooplankton (lower curve) has more intense variability in small-scales than phytoplankton (upper curve), as observed in experiments.}
\label{fig:transect}
\end{figure}

In Fig.~\ref{fig:pattern}, we display the typical phytoplankton and zooplankton patterns that emerge from the model. The patterns presented agree with experimentally observed ones, in which zooplankton display more patchiness than phytoplankton~\cite{Mackas77,Weber86}. It should be pointed out that the patterns generated are not static. In Fig.~\ref{fig:transect}, we present data from a transect of the system, which is analogous to what would be observed by a shipborne sampling method.

\begin{figure}
\begin{center}
\resizebox{0.75\columnwidth}{!}{
 \includegraphics[angle=270,totalheight=3cm]{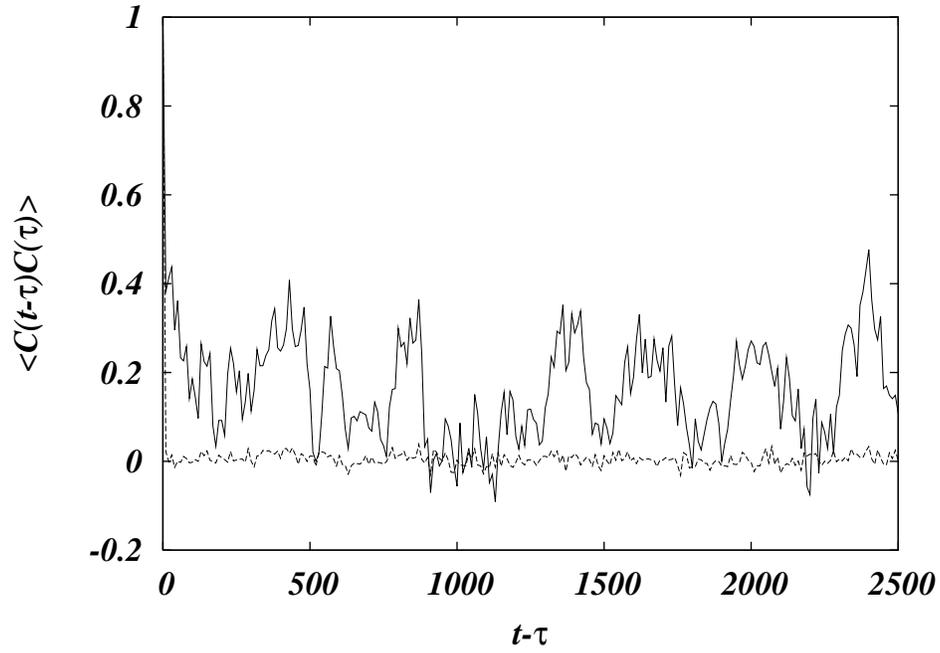} }
\end{center}
\caption{Time correlation functions for phytoplankton and zooplankton obtained after the system is allowed to relax for a long time ($\tau=2500$). After this transient period, the time correlation function of a tracer is equal to $1$, which indicates that the pattern has become stationary (not shown). $C(t)$ represents the concentration of phytoplankton for the upper curve and of zooplankton for the lower curve.}
\label{fig:timecorr}
\end{figure}

\begin{figure}
\begin{center}
\resizebox{0.75\columnwidth}{!}{
 \includegraphics[angle=270,totalheight=3cm]{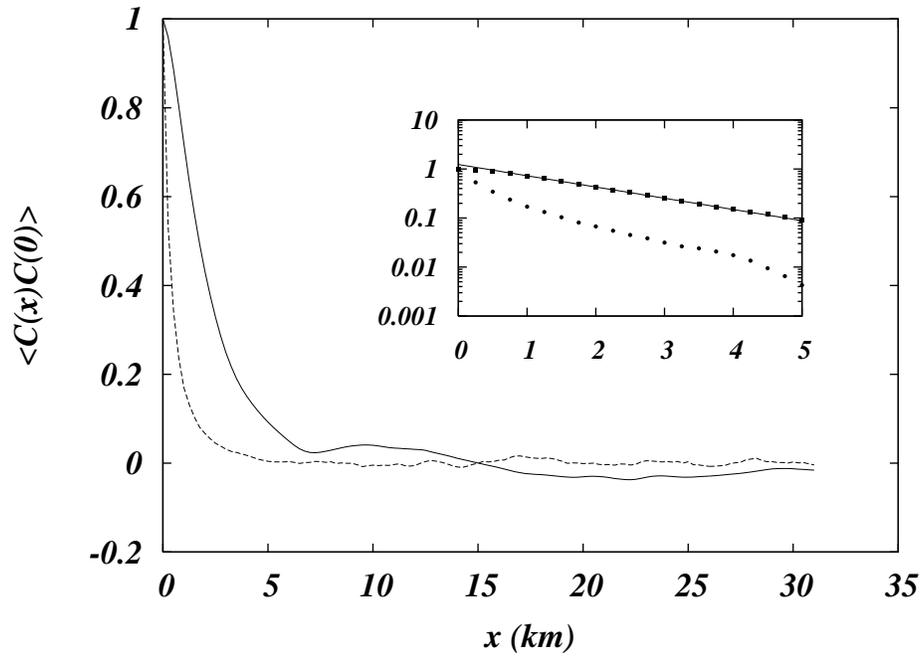} }
\end{center}
\caption{Spatial correlation functions for plankton. $C(x)$ represents the concentration of phytoplankton and of zooplankton.  The zooplankton correlation function (broken line) decays more rapidly than the phytoplankton correlation function (continuous line). Inset: semi-log plot shows that the phytoplankton correlation has an initial exponential decay.}
\label{fig:spacecorr}
\end{figure}

The time correlations of the zooplankton is more rapidly decreasing than that of phytoplankton, since the noise introduced in the zooplankton growth is delta correlated in time. It is expected that the time correlation of zooplankton concentration should decay rapidly to zero. The phytoplankton concentration, on the other hand, is only indirectly affected by this uncorrelated noise, through the nonlinear population dynamics and convection. As such, changing patterns arise, as can be seen by the fluctuating correlation function. Even though there are fluctuations, the average temporal correlation function does not vanish (see Fig.~\ref{fig:timecorr}) and oscillates around a finite value for long times. The phytoplankton patterns generally resemble a tracer pattern, with the difference that the latter becomes static in the long time limit.  Likewise, Fig.~\ref{fig:spacecorr}  displays the spatial correlation functions of the plankton distribution. Again, due to the presence of a noise term that is uncorrelated in space, the zooplankton correlations decay faster than the phytoplankton correlations.   

\section{Spectral analysis}
\label{sec:spectral}

It is common in studies of plankton distribution to compare the power spectra of their distribution with that of physical factors such as temperature. Spectra of sea-surface temperature derived from airborne radiometer measurements have exponents that vary between $-2$ and $-3$. The ones obtained from shipborne measurements vary widely, but are found in a similar range. Gower \emph{et al.}~\cite{Gower80} calculated the power spectrum of phytoplankton concentration patterns averaged over all directions and found that the signal variance is related to the inverse wavelength with an exponent of $-2.92$.  Initially, it was proposed that this exponent, being close to the $-3$ exponent expected for the energy spectrum of geostrophic turbulence (see Fig.~\ref{fig:velocity_spectrum}), indicated that phytoplankton was behaving as a passive tracer. However, it was pointed out that a conserved passive tracer in geostrophic flow should follow a $k^{-1}$ law~\cite{Lesieur81}.  This discrepancy can be understood from the fact that phytoplankton, being microscopic plants, is better described as a reactive tracer because they reproduce, grow, die and are eaten. Besides this, the biological time rates are short when compared to typical time-scales of quasi-geostrophic eddies.  Reactive tracers stirred by geostrophic turbulence generally have spectral slopes in the range from $-1$ to $-3$. Smith \emph{et al.}~\cite{Smith88} found phytoplankton spectral slopes of about $k^{-3}$ offshore, $k^{-2.2}$ inshore and $k^{-1}$ at length scales $<10\,$km by analyzing satellite images. Abraham~\cite{Abraham98,Abraham02} showed that biological factors such as growth rates have a strong influence on the spectral slope. The ratio of the biological and flow timescales is an important factor and  fast-growing organisms such as phytoplankton tend to have steeper spectral slopes.

There have been studies that compare the spectra of plankton patchiness to the $k^{-5/3}$ model of $3$D isotropic turbulence~\cite{Kolmogorov41}. It has been argued in the review by Franks~\cite{Franks05} that planktonic data in these studies do not resolve the appropriate (small) spatial scales in order to make the comparison. Except in intense turbulence, the small spatial scale is characterized by the inertial subrange which lies typically in between scales of about $1$ and $100\,$cm. 

\begin{figure}
\begin{center}
\resizebox{0.75\columnwidth}{!}{
 \includegraphics[angle=270,totalheight=3cm]{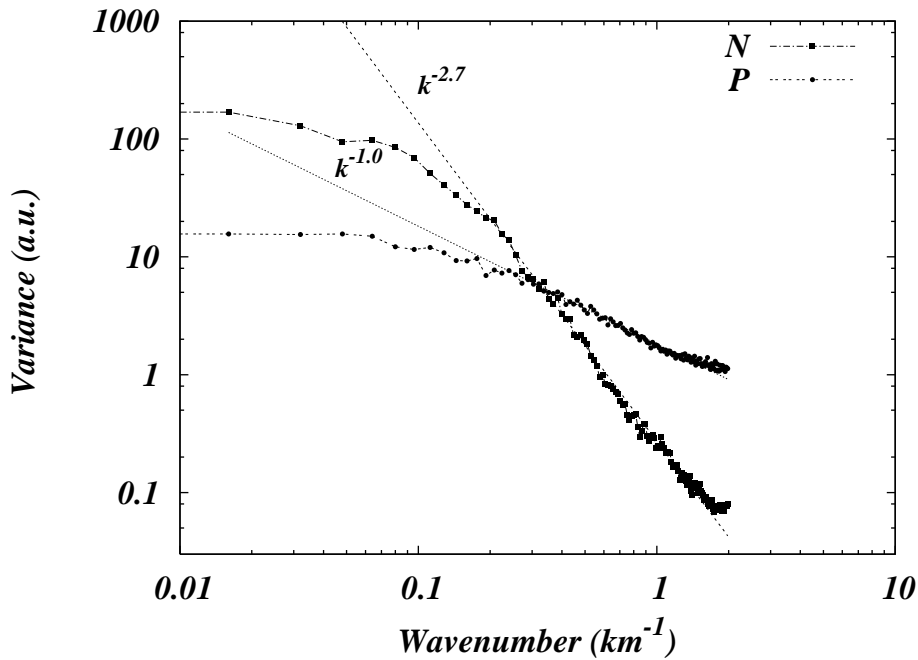} }
\end{center}
\caption{Variance spectra for phytoplankton and zooplankton, obtained from an average over $250$ horizontal and $250$ vertical transects of the system. Phytoplankton displays a steeper slope, as observed in experiments. The exponents obtained are $-2.7 \pm 0.1 $ for phytoplankton and $-1.0 \pm 0.1$ for zooplankton. }
\label{fig:fourier}
\end{figure}

 We analyze larger scales at which the water motion is better described by two-dimensional geostrophic turbulence~\cite{Kraichnan67,Charney71}. In Fig.~\ref{fig:fourier}, we present the plankton spectra obtained from averaging the power spectrum of $500$ transects, $250$ in the vertical direction and $250$ in the horizontal. We obtain spectral slopes comparable to experimental results, with zooplankton having a more white-noise-like spectrum. This is due to the random term in the dynamics that leads to a higher variability in zooplankton. The phytoplankton population is also affected by noise, but in an indirect way. 

\section{Conclusion}
\label{sec:conclusion}

Even though pattern formation in plankton populations is a long known and well documented phenomena, there is still controversy about the mechanisms that give rise to their spatial variability. We believe that simplified models capturing the main features of the population dynamics can help to shed light on the dynamics of the actual system. The addition of noise allows to have a more realistic description, because zooplankton population is smaller and therefore is not as well modelled by a continuous field. Besides this,  the interplay between advection and the growth dynamics with a random term generate results that agree better with experimental data than models based on reaction-diffusion equations.

\acknowledgments
This work was supported by the DGiCYT under grant no. FIS2005-01299, by CNPq and by CAPES. We dedicate this article to the memory of Prof. Carlos P\'erez.

\end{document}